\documentclass [12pt]{article}
\usepackage{amsmath,amssymb,cite}

\usepackage{tabularx}
\usepackage[english]{babel}
\usepackage[dvips]{graphicx}
\usepackage{indentfirst}
\usepackage{epsfig}
\numberwithin{equation}{section}

\begin{document}

\title{
Remarks on the eigenvalues distributions of $D\leq 4$ Yang-Mills matrix models}

\author{Badis Ydri\footnote{Emails:ydri@stp.dias.ie}\\
Institute of Physics, BM Annaba University,\\
BP 12, 23000, Annaba, Algeria.
}

\maketitle
\abstract{The phenomena of emergent fuzzy geometry and noncommutative gauge theory from Yang-Mills matrix models is  briefly reviewed. In particular, the eigenvalues distributions of Yang-Mills matrix models in lower dimensions in the commuting (matrix or Yang-Mills) phase of these models are discussed.
}

\section{Introduction}
A matrix model  of fundamental importance to superstring theory, in particular the BFSS and BMN conjectures \cite{Banks:1996vh,Berenstein:2002jq},  is given by the Euclidean action \cite{Connes:1997cr,Ishibashi:1996xs}:
\begin{eqnarray}
S=\frac{1}{4g^2}Tr_{\cal H}\big(i[{D}_{\mu},{D}_{\nu}]-\theta^{-1}_{\mu\nu}\big)^2+Tr_{\cal H}\bar{\psi}\gamma^{\mu}[{D}_{\mu},\psi].\label{action1}
\end{eqnarray}
Here $D_{\mu}$ are $D$ Hermitian operators in some algebra ${\cal A}$ acting on an abstract (typically infinite dimensional) Hilbert space ${\cal H}$. The trace $Tr_{\cal H}$ carries the dimension of  $({\rm length})$,  the non-commutativity tensor ${\theta}$ has the dimension of $({\rm length})^2$, the connection operators ${D}_{\mu}$ have dimension of $({\rm length})^{-1}$ and the coupling constant $g$ is of dimension $(\rm mass)^{2-{D}/{2}}$.  The gamma matrices are ${\cal N}\times {\cal N}$ matrices with ${\cal N}=2(D-2)$ and therefore the spinor $\psi$ is ${\cal N}$ dimensional, where every component is an (odd) element in the algebra  ${\cal A}$ acting  on the Hilbert space ${\cal H}$. The Majorana condition reads $\bar{\psi}=\psi^TC$ where $C$ is the charge conjugation matrix. The model has an obvious gauge symmetry $D_i\longrightarrow U^+D_iU$, $\psi\longrightarrow U^+\psi U$, and it is supersymmetric in dimensions $D=3,4,6,10$. 

In the case of the algebra of complex $N\times N$ matrices, i.e. ${\cal A}=Mat_N({\bf C})$, the above model corresponds precisely to the reduction of the $D-$dimensional $U(N)$ Yang-Mills theory to zero dimension, which is known to be relevant to the physics of $D0$ branes \cite{Witten:1995im}. In particular, the $D=10$ model is the IKKT matrix model proposed in \cite{Ishibashi:1996xs} as a non-perturbative definition of type IIB superstring theory. The corresponding partition function exists (for sufficiently large $N$) only in dimensions $D=4,6,10$ \cite{Austing:2001pk}, whereas the quenched approximations exist in $D=3,4,6,10$ \cite{Krauth:1998yu}.

This model provides a background independent formulation of non-commutative gauge theories \cite{Aoki:1999vr,Seiberg:2000zk,Polychronakos:2000zm}, with exact supersymmetry. Indeed, non-commutative geometry \cite{Connes:1994yd} is the only known non-trivial extension of supersymmetry as pointed out in \cite{Douglas:2001ba}. 

Obviously the global minimum of the model (\ref{action1}) are connection operators ${B}_{\mu}=\theta^{-1}_{\mu\nu}\hat{x}_{\nu}$, where $\hat{x}_{\mu}$ are operators satisfying the Heisenberg algebra 
\begin{eqnarray}
[\hat{x}_{\mu},\hat{x}_{\nu}]=\theta^{-1}_{\mu\nu}.\label{eom1} 
\end{eqnarray}
The operators $\hat{x}_{\mu}$ can be identified with the coordinate operators on the non-commutative Moyal-Weyl space ${\bf R}^D_{\theta}$, while derivations are given by $\hat{\partial}_{\mu}=-iB_{\mu}$. The sector of this matrix theory which corresponds to a non-commutative $U(n)$ gauge field on Moyal-Weyl space ${\bf R}^D_{\theta}$ is obtained by expanding ${D}_{\mu}$ around ${B}_{\mu}$ as $D_{\mu}=B_{\mu}\otimes {\bf 1}_n+A_{\mu}$, and as a consequence, the curvature becomes $F_{\mu\nu}=-i[{D}_{\mu},{D}_{\nu}]+\theta^{-1}_{\mu\nu}=[\hat{\partial}_{\mu},A_{\nu}]-[\hat{\partial}_{\nu},A_{\mu}]-i[A_{\mu},A_{\nu}]$. Indeed, by employing the Weyl map \cite{weyl:1931} from operators to fields, and the Moyal-Weyl star product \cite{Groenewold:1946kp,Moyal:1949sk}, we can rewrite the resulting action as a $U_*(n)$ gauge theory on ${\bf R}^D_{\theta}$ of the type found in \cite{Seiberg:1999vs} in low energy limit of string theory. More precisely, we obtain the action
\begin{eqnarray}
S=\frac{1}{4g^2}\int d^dx \big({\partial}_{\mu} A_{\nu}-\hat{\partial}_{\nu} A_{\mu}-i[A_{\mu},A_{\nu}]_*\big)^2+\int d^dx \bar{\psi}\gamma^{\mu}\big(i\partial_{\mu} \psi+[A_{\mu},\psi]_*\big).\label{action2}
\end{eqnarray}
It is well established that there are no finite dimensional representations of the Heisenberg algebra (\ref{eom1}). However, we can still formulate finite dimensional truncation of the matrix model (\ref{action1}), based on the non-commutative fuzzy torus \cite{Ambjorn:2000cs,Ambjorn:1999ts}, which relies on a twisted Eguchi-Kawai reduction \cite{Eguchi:1982nm}.  Some remarkable results using this non-perturbative regularization can be found for example in \cite{Bietenholz:2006cz,Azeyanagi:2007su}.

In this note we are mainly interested in a non-perturbative regularization of the above matrix model for $d=2$ (and thus $\theta_{ij}=\theta\epsilon_{ij}$)  using $N\times N$ matrix models on the fuzzy sphere \cite{Hoppe:1982,Madore:1991bw}. Constructions along this direction can be found for example in \cite{CarowWatamura:1998jn,Presnajder:2003ak,Steinacker:2003sd,Steinacker:2007iq}. However, the starting point here is the result of \cite{Alekseev:2000fd} that the dynamics of open strings moving in a curved space with ${\bf S}^3$ metric, in the presence of a non-vanishing Neveu-Schwarz $B-$field, and with $Dp-$branes, is equivalent to leading order in the string tension to a gauge theory on a non-commutative fuzzy sphere with a Myers (Chern-Simons) term \cite{Myers:1999ps}. Their action is
\begin{eqnarray}
S=N Tr\big(-\frac{1}{4}[X_a,X_b]^2+\frac{2i\alpha}{3}\epsilon_{abc} X_aX_bX_c\big).\label{action30}
\end{eqnarray}
This is a $D=3$ Yang-Mills matrix model with a Chern-Simons term. Yang-Mills matrix models, including various mass terms, and with and without supersymmetry, in dimension $3$, were studied extensively in recent years using both the Monte Carlo method and various other powerful analytical tools. See for example \cite{Ydri:2012bq} for an extensive list of references. This interest is due mainly, as discussed above, to their relation to string theory on ${\bf S}^3$, to noncommutative geometry and to the fuzzy sphere. The Yang-Mills term provides a simpler quenched analogue of the celebrated IKKT model\cite{Ishibashi:1996xs}, whereas the Yang-Mills model with a Chern-Simons term contains the fuzzy sphere \cite{Hoppe:1982,Madore:1991bw} as a global minimum for some range of the parameters. The addition of other mass terms and/or Majorana fermions define a generic physics which interpolates between these two cases. This later model has a rich and subtle phase structure which is discussed, for example, in the recent work \cite{O'Connor:2013rla}, but also it is intimately related to Yang-Mills matrix models in dimension $2$ and dimension $4$. 

In these notes, we will largely focus on the so-called matrix or Yang-Mills phase, in which the matrices are nearly commuting, and attempt to provide a synthesis of the different construction and limits in which the parabolic eigenvalues distributions, observed in this phase, can be derived. 

This article is organized as follows:
\begin{itemize}
\item Section $2$: The $D\leq 4$ Yang-Mills matrix models.
\begin{itemize}
\item The basic $D=3$ action.
\item The commutative limit and the Chern-Simons action.
\item $4-$dimensional extension and phase structure.
\item Supersymmetric extension.
\item Matrix model solution.
\item But where is the fuzzy sphere?
\end{itemize}
\item Section $3$: Synthesis of other approaches.
\begin{itemize}
\item Hoppe and inverted oscillator problems.
\item The three-color problem.
\end{itemize}
\item Section $4$: Conclusion and Outlook.
\item Appenidx: Supersymmetry and Localization.
\end{itemize}

\section{The $D\leq 4$ Yang-Mills matrix models}
\subsection{The basic $D=3$ action}
The low energy dynamics of open strings moving in a background magnetic field with ${\bf S}^3$ metric is described by the three-matrix model \cite{Alekseev:1999bs,Alekseev:2000fd,Hikida:2001py}
\begin{eqnarray}
S_0={\rm YM}+{\rm CS}=-\frac{N}{4}Tr[X_a,X_b]^2+\frac{2iN\alpha}{3}{\epsilon}_{abc}TrX_aX_bX_c.\label{fund}
\end{eqnarray}
The $X_a$, $a=1,2,3$, are three $N\times N$ Hermitian matrices, $c_2=(N^2-1)/4$ is the quadratic Casimir of $SU(2)$ in the irreducible representation $(N-1)/2$, and $\alpha$ is the parameter of the model which is related to the gauge coupling constant $g$ and to the inverse temperature $\beta$ by
\begin{eqnarray}
\tilde{\alpha}=\alpha\sqrt{N}~,~\tilde{\alpha}^4=\frac{1}{g^2}=\beta.
\end{eqnarray}
This theory consists of the Yang-Mills term ${\rm YM}$, which can be obtained from the reduction to zero dimension of ordinary three-dimensional $U(N)$ Yang-Mills theory, and a Chern-Simons term ${\rm CS}$ due to Myers effect \cite{Myers:1999ps}. This model was also introduced in \cite{Iso:2001mg,CarowWatamura:1998jn} as a noncommutative gauge theory on the fuzzy sphere. It contains, beside the usual two-dimensional gauge field, a scalar fluctuation normal  to the sphere which can be given by \cite{Karabali:2001te}
\begin{eqnarray}
\Phi=\frac{X_a^2-{\alpha}^2c_2}{2\alpha^2\sqrt{c_2}}.
\end{eqnarray} 
The Yang-Mills and Chern-Simons actions in (\ref{fund}) are given explicitly by
\begin{eqnarray}
{\rm YM}=\frac{N}{4}Tr\bigg(i[X_a,X_b]+\epsilon_{abc}\alpha X_c\bigg)^2.
\end{eqnarray}
\begin{eqnarray}
{\rm CS}=-\frac{N\alpha}{6}\epsilon_{abc}Tr\bigg(i[X_a,X_b]+\alpha \epsilon_{abd} X_d\bigg)X_c-\frac{N\alpha^2}{6}TrX_a^2.\label{CS}
\end{eqnarray}
The action $S_0$ is invariant under the unitary transformations $X_a\longrightarrow UX_aU^+$ where $U\in U(N)$, as well as under global $SO(3)$ rotations. It enjoyes also the extra symmetry $X_a\longrightarrow X_a+\alpha_a{\bf 1}_N$, and as a consequence, we can choose $X_a$ to be traceless, viz $Tr X_a=0$. The equations of motion are
\begin{eqnarray}
[X_c,F_{ab}]=0~,~F_{ab}=\frac{1}{\alpha^2}\bigg(i[X_a,X_b]+\alpha \epsilon_{abd} X_d\bigg).
\end{eqnarray}
Extrema of the models are given by $i)$ reducible representation of $SU(2)$, and $ii)$ commuting matrices. The classical absolute minima of the model of the model is given by the irreducible representation of $SU(2)$ of dimension $N$, viz $X_a=\alpha L_a$, where $[L_a,L_b]=i\epsilon_{abc}L_c$ and $L_a^2=c_2$. Small fluctuations around this background are seen to have the geometrical content of a $U(1)$ gauge field coupled to a scalar field on a background fuzzy sphere. Indeed, by writing $X_a=\alpha(L_a+A_a)$, we find $F_{ab}=i[L_a,A_b]-i[L_b,A_a]+\epsilon_{abc}A_c+i[A_a,A_b]$ and $\Phi=({x}_aA_a+A_a{x}_a+A_a^2/\sqrt{c_2})/2$,   where ${x}_a$ are the coordinate operators on the fuzzy sphere defined by ${x}_a=L_a/\sqrt{c_2}$. The   Yang-Mills and Chern-Simons actions take the equivalent form (with $F_{ab}^{(0)}=i[L_a,A_b]-i[L_b,A_a]+\epsilon_{abc}A_c$)
\begin{eqnarray}
{\rm YM}=\frac{1}{4g^2}Tr F_{ab}^2.
\end{eqnarray}
\begin{eqnarray}
{\rm CS}=-\frac{1}{2g^2}\epsilon_{abc}Tr \bigg(\frac{1}{2}F_{ab}^{(0)}A_c+\frac{i}{3}[A_a,A_b]A_c\bigg).
\end{eqnarray}
\subsection{The commutative limit and the Chern-Simons action}
In the commutative limit $N\longrightarrow \infty$, we use the coherent states and star product on the fuzzy sphere \cite{Balachandran:2001dd}. In this limit, we can also divide the gauge field $A_a$ into a tangential gauge field $a_a$ and a normal component $n_a\phi$, i.e. $A_a=a_a+n_a\phi$, where $n_an_a=1$ and $n_a\phi=0$. Obviously, in this limit we have ${x}_a\longrightarrow n_a$, and hence $\Phi=n_aA_a=\phi$. Furthermore, both $F_{ab}$ and $F_{ab}^{(0)}$ tend, in this limit, to $F_{ab}=f_{ab}+(i{\cal L}_a\phi)n_b-(i{\cal L}_b\phi)n_a-\epsilon_{abc}n_c\phi$, where $f_{ab}=i{\cal L}_aa_b-i{\cal L}_ba_a+\epsilon_{abc}a_c$. We have then the limits 
\begin{eqnarray}
{\rm YM}=\frac{1}{4g^2}\int_{{\bf S}^2}\frac{d\Omega_2}{4\pi}\bigg(f_{ab}^2-2\epsilon_{abc}n_cf_{ab}\phi-2({\cal L}_a\phi)^2+2\phi^2\bigg).\label{YMNinfty}
\end{eqnarray}
\begin{eqnarray}
{\rm CS}=\frac{1}{4g^2}\int_{{\bf S}^2}\frac{d\Omega_2}{4\pi}\bigg(-2\epsilon_{abc}n_cf_{ab}\phi+2\phi^2\bigg).\label{CSNinfty}
\end{eqnarray}
The commutative limit of the action $S_0$ is given by the sum of the above two terms. As one can immediately
see, this theory consists of a $2$-component gauge field $a_a$ that 
mixes with a scalar field $\phi$. The presence of the scalar field means that the geometry
is completely specified, in that all the ingredients of the spectral triple
are supplied by this field. In contrast a two-dimensional gauge theory on its
own would not be sufficient to specify the geometry.

In order to see this more clearly, we introduce $\partial_{\mu}$ and $a_{\mu}$, with $\mu=\theta,\phi$,  by the relations ${\cal L}_a=L_a^{\mu}\partial_{\mu}$ and $a_a=L_a^{\mu}a_{\mu}$. Using ${\cal L}_a=-i\epsilon_{abc}n_b\partial_c$, we have explicitly  $L_1^{\theta}=i\sin\phi$, $L_1^{\phi}=i\cot\theta\cos\phi$, $L_1^{\theta}=-i\cos\phi$, $L_1^{\phi}=i\cot\theta\sin\phi$, $L_3^{\theta}=0$, $L_3^{\phi}=-i$. The metric $g_{\mu\nu}$ on the sphere is $ds^2=d\theta^2+\sin^2\theta d\phi^2$, while the inverse metric $g^{\mu\nu}$ can be expressed as $g^{\mu\nu}=-L_a^{\mu}L_a^{\nu}$. We can also verify the important identities $i\epsilon_{abc}L_c^{\nu}=L_a^{\mu}\partial_{\mu}L_b^{\nu}-L_b^{\mu}\partial_{\mu}L_a^{\nu}$, and $\epsilon_{abc}n_cL_a^{\mu}L_b^{\nu}=-\epsilon^{\mu\nu}$, where $\epsilon^{\theta\phi}=1/\sin\theta$. We compute then $f_{ab}=iL_a^{\mu}L_b^{\nu}f_{\mu\nu}$, $\epsilon_{abc}n_cf_{ab}=-i\epsilon^{\mu\nu}f_{\mu\nu}$, where  $f_{\mu\nu}=\partial_{\mu}a_{\nu}-\partial_{\nu}a_{\mu}$, and $f_{ab}^2=-f_{\mu\nu}f^{\mu\nu}$, $(\epsilon_{abc}n_cf_{ab})^2=-2f^{\mu\nu}f_{\mu\nu}$.  Thus, by integrating $\phi$ in (\ref{YMNinfty})+(\ref{CSNinfty}), we obtain the gauge theory on the sphere given by the action 
\begin{eqnarray}
{\rm YM}+{\rm CS}&=&\frac{1}{4g^2}\int_{{\bf S}^2}\frac{d\Omega_2}{4\pi}\bigg(f_{ab}^2-2(\epsilon_{abc}n_cf_{ab})\frac{1}{{\cal L}_a^2+2}(\epsilon_{abc}n_cf_{ab})\bigg)\nonumber\\
&=&\frac{1}{4g^2}\int_{{\bf S}^2}\frac{d\Omega_2}{4\pi}\bigg(-f_{\mu\nu}f^{\mu\nu}+2(\epsilon_{\mu\nu}f_{\mu\nu})\frac{1}{-\partial^{\mu}\partial_{\mu}+2}(\epsilon_{\mu\nu}f^{\mu\nu})\bigg).\label{CL}
\end{eqnarray}
The second term can be canceled by making the mass of the normal scalar field $\phi$ sufficiently large. This can be achieved by adding to the action $S_0$ a potential $V$ given by \cite{CastroVillarreal:2004vh}
\begin{eqnarray}
V=\frac{m^2N}{2c_2}Tr (X_a^2-\alpha^2c_2)^2.\label{fundV}
\end{eqnarray}
 It is obvious, from (\ref{CSNinfty}), that the matrix Chern-Simons action (\ref{CS}) leads also to a gauge theory on the sphere. However the commutative limits of (\ref{fund}) and (\ref{CS}) are not the same. Indeed, by integrating $\phi$ in (\ref{CSNinfty}), we obtain a  gauge theory on the sphere given by the action 
\begin{eqnarray}
{\rm CS}&=&-\frac{1}{8g^2}\int_{{\bf S}^2}\frac{d\Omega_2}{4\pi}\big(\epsilon_{abc}n_cf_{ab}\big)^2\nonumber\\
&=&\frac{1}{4g^2}\int_{{\bf S}^2}\frac{d\Omega_2}{4\pi}f^{\mu\nu}f_{\mu\nu}.
\end{eqnarray}
This is the canonical gauge theory on the sphere. 

Furthermore, the equations of motion arising from the Chern-Simons action (\ref{CS}) read $[X_a,X_b]=i\alpha\epsilon_{abc}X_c$, and as a consequence, the fuzzy sphere configurations $X_a=\alpha L_a$ are still solutions. In other words, it is expected that the model (\ref{CS}) contains all the essential features of the phenomena of emergent geometry observed in Monte Carlo simulations of $S_0$.  Indeed, the condensation of a background spherical geometry at low temperatures $1/\beta$, and the appearance of the phase of commuting matrices at high temperatures, are two  effects sustained, as we will discuss further shortly, in the  Chern-Simons matrix model (\ref{CS}).

Hence, we propose here to consider the simpler matrix model given only by the Chern-Simons action ${\rm CS}$, modulo an arbitrary mass term for $X_a$, namely 
\begin{eqnarray}
S_{\rm CS}=-{\rm CS}+\frac{N\alpha^2\tau}{2}Tr X_a^2=\frac{N\alpha}{2}Tr \bigg(\alpha (1+\tau)X_a^2+\frac{i}{3}\epsilon_{abc}[X_a,X_b]X_c\bigg).\label{CS1}
\end{eqnarray}
This action has the same commutative limit as the model with $\tau=0$. In this article, we will construct explicitly, starting from a mass deformed $D=4$ supersymmetric Yang-Mills matrix model, a $D=3$ bosonic Yang-Mills matrix model, analogous to $S_0$,  which is quantum mechanically  equivalent to a Chern-Simons matrix model.   

By integrating over the matrix $X_3$, then performing the scaling $X_i\longrightarrow \alpha(1+\tau)D_i$, and defining $t=\alpha^4/2(1+\tau)^3$, we obtain the effective path integral
\begin{eqnarray}
Z_{\rm eff}=\int dD_1 dD_2 \exp\bigg(-Nt Tr[D_1,D_2]^2-Nt TrD_i^2\bigg).\label{eff0}
\end{eqnarray}
As noted in \cite{Ishiki:2008vf}, the above matrix model (\ref{CS1}) may be regarded  as the ${\cal N}=1^*$ Dijkgraaf-Vafa  theory, i.e. as a mass deformed superpotential of ${\cal N}=4$ supersymmetric Yang-Mills theory \cite{Dijkgraaf:2002fc,Dijkgraaf:2002vw,Dijkgraaf:2002dh}. A more rigorous derivation of gauge theory on the sphere starting from the above matrix model is also given in  \cite{Ishiki:2008vf,Ishiki:2009vr}.

\subsection{$4-$dimensional extension and phase structure}
The extension of (\ref{fund}) to four dimensions is straightforward given by
\begin{eqnarray}
S_1=-\frac{N}{4}Tr[X_{\mu},X_{\nu}]^2+\frac{2iN\alpha}{3}{\epsilon}_{abc}TrX_aX_bX_c.\label{fund1}
\end{eqnarray}
This is a four-matrix model. This model suffers from the same phase transition as the original model (\ref{fund}).  The emergent geometry transition to the  fuzzy sphere occurs in the two cases at the values
\begin{eqnarray}
\tilde{\alpha}_*=2.1\pm 0.1~~{\rm for}~S_0.\label{lcritical1}
\end{eqnarray}
\begin{eqnarray}
\tilde{\alpha}_*=2.55\pm 0.1~~{\rm for}~S_1.\label{lcritical2}
\end{eqnarray}
Monte Carlo studies of the $D=3$ Yang-Mills matrix model $S_0$ is found in \cite{Azuma:2004zq,O'Connor:2006wv,DelgadilloBlando:2008vi,DelgadilloBlando:2007vx}. More recent studies are found in  \cite{DelgadilloBlando:2012xg,O'Connor:2013rla}.  The $D=4$ model $S_1$ is studied in  \cite{Ydri:2012bq}.

The transition from/to the fuzzy sphere phase was found to have a discontinuity in the internal energy, i.e. the transition is associated with a non-zero latent heat. The corresponding specific heat  diverges at the transition point from the sphere side, while it remains constant from the matrix side. This indicates a second order behaviour with critical fluctuations only from the sphere side. Furthermore, we observe a discontinuity in the order parameter indicating that the transition is first order. The order parameter is identified with the radius of the sphere defined by
\begin{eqnarray}
R=\frac{1}{\tilde{\alpha}^2c_2}<Tr X_a^2>.
\end{eqnarray}
The ground state configurations, in the fuzzy sphere phase, are given by
\begin{eqnarray}
X_a=\alpha L_a~,~{\rm for}~S_0~~~;~~~X_4=0~,~X_a=\alpha L_a~,~{\rm for}~S_1.\label{b1}
\end{eqnarray}
In other words we have a fuzzy spherical geometry given by the commutation relations $[X_4,X_a]=0$, $[X_a,X_b]=i\epsilon_{abc}\alpha X_c$.

In the matrix phase, it was found that the joint eigenvalues distribution of the matrices $X_1$, $X_2$,...$X_D$ is uniform inside a solid ball of some radius $L$. These distributions are given explicitly by

\begin{eqnarray}
\rho(x)=\frac{3}{4L^3}(L^2-x^2)~,~L=2~,~{\rm for}~S_0.\label{3ball}
\end{eqnarray}

\begin{eqnarray}
\rho(x)=\frac{8}{3\pi L^4}(L^2-x^2)^{\frac{3}{2}}~,~L=1.83~,~{\rm for}~S_1.\label{4ball}
\end{eqnarray}
In summary, the two different phases of the Yang-Mills matrix models $S_0$ and $S_1$ are characterized by 
\begin{center}
\begin{tabular}{|c|c|}
\hline
fuzzy sphere ($\tilde{\alpha}>\tilde{\alpha}_*$ )& matrix phase ($\tilde{\alpha}<\tilde{\alpha}_*$)\\
\hline
$R=1$ & $
R=\frac{1}{2c_2}\frac{L^2}{\alpha^2}$~,~{\rm for}~$S_0$ \\
$R=1$ & $
R=\frac{3}{5c_2}\frac{L^2}{\alpha^2}$~,~{\rm for}~$S_1$ \\
\hline
$C_v=1$  & $C_v=0.75$~,~{\rm for}~$S_0$  \\
$C_v=1.5$ & $C_v=1$~,~{\rm for}~$S_1$\\
\hline
\end{tabular}
\end{center}

\subsection{Supersymmetric extension}
The primary goal, in this article, is to solve analytically the model $S_0$ given by equation  (\ref{fund}). Unfortunatley, this can not be done directly, and an escalation of the problem to supersymmetric massive Yang-Mills in $4$ dimensions is required. Going to $4$ dimensions is essential because supersymmetric Yang-Mills path integrals are not convergent in $D=3$ but are convergent in $D=4$  \cite{Austing:2001pk}. 

The  mass deformed supersymmetric $D=4$ Yang-Mills matrix model, we will consider below, will be dominated by  saddle points in which the four bosonic matrices $X_{\mu}$  are constrained in such a way that only three of them are effectively independent, while the fermionic matrices become frozen or decoupled. Indeed, the resulting  model can be reduced, by means of localization technique, to the matrix Chern-Simons action  (\ref{CS1}), which actually yields  the $3-$dimensional eigenvalues distribution (\ref{3ball}). This mass deformed supersymmetric $D=4$ Yang-Mills matrix model is therefore  effectively analogous  to the bosonic $3-$dimensional Yang-Mills matrix model $S_0$.   We summarize here these results and leave the detailed derivation for the next section.

It is clear that inclusion of supersymmetry is more subtle due to the Myers term. We will consider, in the following, the two approaches of mass  and cohomological deformations. The resulting action is a one-parameter matrix model given by equation (\ref{fund2F1}). By performing the scaling $X_{\mu}\longrightarrow N^{1/4}X_{\mu}$, and defining $-\kappa_1=N^{1/4}\alpha$, we get
\begin{eqnarray}
S_2&=&-\frac{N}{4}Tr[X_{\mu},X_{\nu}]^2+\frac{2iN\alpha}{3}\epsilon_{abc}TrX_aX_bX_c+\frac{N\alpha^2}{4}Tr X_a^2-\frac{N\alpha^2}{4}TrX_4^2\nonumber\\
&-&Tr{\theta}^+\bigg(i[X_4,..]+{\sigma}_a[X_a,..]+\alpha\bigg)\theta.\label{fund2}
\end{eqnarray}
The mass deformed and the cohomologically deformed supersymmetric Yang-Mills matrix models studied in   \cite{Ydri:2012bq}  and \cite{Anagnostopoulos:2005cy}, mainly by means of Monte Carlo, are closely related to the model (\ref{fund2}). These models should be viewed as $4-$dimensional mass deformed analogues of the IIB (IKKT) matrix model  \cite{Ishibashi:1996xs}. The supersymmetric Yang-Mills matrix model (\ref{fund2}) is, on the other hand,  very special as it is effectively equivalent to the bosonic $3-$dimensional Yang-Mills matrix model $S_0$.

We note that the background configurations of interest here are
\begin{eqnarray}
X_4=0~,~X_a=\varphi \alpha L_a~,~\varphi=\frac{1}{2}.\label{b2}
\end{eqnarray}
The action $S_2$ given by (\ref{fund2}) admits the same commutative limit as $S_0$, i.e. (\ref{CL}), because the bosonic matrix $X_4$ and the fermionic matrices $\theta^+$ and $\theta$ become free decoupled fields  in this limit \footnote{We note that only the $U(1)$ gauge group can be realized, i.e. is stable, in these matrix models.}. By employing supersymmetry and localization technique,   we can show that this theory is equivalent to the  matrix model
\begin{eqnarray}
\tilde{S}_{\rm CS}=-2N\alpha TrX_4[X_1,X_2]-\frac{N\alpha^2}{2}TrX_4^2+\frac{N\alpha^2}{4}TrX_i^2.\label{CSfund0}
\end{eqnarray}
The dominant saddle point in the direction of the BRST field $\bar{\phi}=-(X_3-iX_4)/2$ was found to be given by $\bar{\phi}=0$ or equivalently $X_3=iX_4$, and as a consequence, the above action can be formally regarded as the Chern-Simons matrix action  
\begin{eqnarray}
\tilde{S}_{\rm CS}=2iN\alpha TrX_3[X_1,X_2]+\frac{N\alpha^2}{2}TrX_3^2+\frac{N\alpha^2}{4}TrX_i^2.\label{CSfund}
\end{eqnarray}
This should be compared with (\ref{CS1}). By integrating over $X_3$, then performing the scaling $X_i\longrightarrow \alpha D_i/\sqrt{8}$, and defining $t=\alpha^4/32$,  we obtain the effective path integral
\begin{eqnarray}
Z_{\rm eff}=\int dD_1 dD_2 \exp\bigg(-Nt Tr[D_1,D_2]^2-Nt TrD_i^2\bigg).\label{eff1}
\end{eqnarray}
This should be compared with (\ref{eff0}) with the identification 
\begin{eqnarray}
\frac{\alpha}{(1+\tau)^{3/4}}\leftrightarrow \frac{\alpha}{2} 
\end{eqnarray}
The analytic continuation of this model is essentially the model studied in  \cite{Kazakov:1998ji}. However, the explicit solution constructed here is quite different from the sophisticated implicit solution found in  \cite{Kazakov:1998ji}. The eigenvalues distribution, we will derive shortly, is also different from the one presented in  \cite{Berenstein:2008eg}, and it agrees very well with Monte Carlo simulations.

\subsection{Matrix model solution} 

We start the solution by diagonalizing the matrix $D_1$ by writing $D_1=U\Lambda U^+$, where $U\in U(N)$ and $\Lambda={\rm diag}(\lambda_1,...,\lambda_N)$, and then integrating over the matrix $D_2$. We obtain the eigenvalues problem   \cite{Eynard:1998fn, Ydri:2012bq}
\begin{eqnarray}
Z_{\rm eff}
&=&\int \prod_{i=1}^Nd\lambda_i 
\prod_{i<j}(\lambda_i-\lambda_j)^2 \prod_{i<j}\big(-Nt(\lambda_i-\lambda_j)^2+Nt\big)^{-1} \exp\big(-Nt\sum_{i}\lambda_i^2\big).\label{patheff0}\nonumber\\
\end{eqnarray}
The effective potential derived from the path integral (\ref{patheff0}) is given by
\begin{eqnarray}
-V_{\rm eff}(\lambda_i)
&=&-Nt\sum_{i}\lambda_i^2+\frac{1}{2}\sum_{i\neq j}\ln (\lambda_i-\lambda_j)^2-\frac{1}{2}\sum_{i\neq j}\ln\big(-t(\lambda_i-\lambda_j)^2+t\big).\nonumber\\
\end{eqnarray}
The saddle point associated with this potential is essentially the inverted oscillator problem which is the analytic continuation of the supersymmetric model considered in   \cite{Moore:1998et}. This saddle point is also related to the Baxter's three-colorings problem \cite{baxter}.

The saddle point equation reads explicitly
\begin{eqnarray}
2 t\lambda_k
&=&\frac{2}{N}\sum_{i\neq k}\frac{1}{\lambda_k-\lambda_i}-\frac{1}{N}\sum_{j\neq k}\bigg[\frac{1}{1+\lambda_k-\lambda_j}-\frac{1}{1-\lambda_k+\lambda_j}\bigg].\label{eigenvalueP}
\end{eqnarray}
We introduce the density of eigenvalues given by
\begin{eqnarray}
\rho(x)=\frac{1}{N}\sum_i\delta(x-\lambda_i)~,~\int dx \rho(x)=1.
\end{eqnarray}
The saddle point equation becomes
\begin{eqnarray}
t z=\int dy \frac{\rho(y)}{(z-y)(1-(z-y)^2)}.\label{sad5}
\end{eqnarray}
Let us assume a wide one-cut support $[-L,L]$. The saddle point equation to leading contribution in $1/L$ takes then the form
\begin{eqnarray}
t z
&=&\rho(z)\big[-\frac{2z}{L^3}-\frac{2z(2z^2+1)}{L^5}-...\big]-\rho^{'}(z)\big[\frac{2}{L}+\frac{2(1+3z^2)}{3L^3}+...\big]\nonumber\\
&+&\rho^{''}(z)\big[-\frac{z}{L}-\frac{z(z^2+3)}{3L^3}-...\big]+...
\end{eqnarray}
Let us assume a quadratic distribution $\rho(z)=a+b z^2$.  The saddle point equation takes then the form
\begin{eqnarray}
t z
&=&z\bigg[a\bigg(-\frac{2}{L^3}-\frac{2}{L^5}+O(7)\bigg)+2b\bigg(-\frac{3}{L}-\frac{5}{3L^3}+O(5)\bigg)\bigg]\nonumber\\
&+&z^3\bigg[a\bigg(-\frac{4}{L^5}+O(7)\bigg)+2b\bigg(-\frac{10}{3L^3}+O(5)\bigg)\bigg]\nonumber\\
&+&...
\end{eqnarray}
The density of eigenvalues $\rho(x)$ must be normalized which gives  immediately $a=(3-2bL^3)/6L$. By assuming also that $\rho(\pm L)=0$, we get

\begin{eqnarray}
\rho(z)=\frac{3}{4L^3}(L^2-z^2).\label{dist}
\end{eqnarray}
By substituting $a=3/4L$ and $b=-3/4L^3$ into the above saddle point equation, we obtain the equation
\begin{eqnarray}
t z
&=&z\bigg[\frac{3}{L^4}+\frac{1}{L^6}+O(8)\bigg]+z^3\bigg[\frac{2}{L^6}+O(8)\bigg]+...
\end{eqnarray}
We obtain immediately the prediction 
\begin{eqnarray}
t=\frac{3}{L^4}\Leftrightarrow L=\frac{96^{1/4}}{\alpha}.\label{scaling3color}
\end{eqnarray}
Let us recall that $z$ are the eigenvalues of $D_1$. The eigenvalues of $X_1=\alpha D_1/\sqrt{8}$ are then given by $z_0=\alpha z/\sqrt{8}$. Furthermore, by comparing (\ref{b1}) and (\ref{b2}) we see that we must also scale $z_0$ as $z_0\longrightarrow 2z_0$. with an eigenvalues distribution given by
  \begin{eqnarray}
\rho_0(z_0)=\frac{3}{4L_0^3}(L_0^2-z_0^2).
\end{eqnarray}
The length of the support $L_0$ is given by
  \begin{eqnarray}
L_0=\frac{2\alpha}{\sqrt{8}}L=2(\frac{3}{2})^{1/4}=2.21.\label{res}
\end{eqnarray}
This is independent of  $\alpha$ which is precisely what we observe in Monte Carlo simulations. We note that  our original estimation of the value of $L_0$ reported in \cite{Ydri:2012bq} contained an error.


Since the above distribution  will only work for large $L$, it must only be valid for small $\alpha$. This is the regime of the matrix phase. We must have therefore the following lower estimate of the critical value
\begin{eqnarray}
  L>>1\Leftrightarrow  \frac{\alpha}{2}<<\frac{\alpha_*}{2}=1.57.
\end{eqnarray}
This may be compared with the ''upper'' critical point observed  for the bosonic model  (\ref{fund}) in \cite{Azuma:2004zq}. This is different from the ''lower'' critical point (\ref{lcritical1}), which indicates a first order transition, and a hysteresis effect. All this may suggest that the fuzzy sphere phase is in fact a metastable state. This is also indicated, somewhat, by the recent Monte Carlo results of \cite{O'Connor:2013rla}.

The above distribution (\ref{dist}) is the same distribution found for small values of $\alpha$ for the corresponding antihermitian model in \cite{Berenstein:2008eg}. However, the crucial difference is the functional dependence of $L$ on $\alpha$. Indeed, they found the behavior $L\sim 1/\alpha ^{4/3}$ which simply does not agree with numerical results. More on this in the next section.

\subsection{But where is the fuzzy sphere?}
We would like also to discuss how does the fuzzy sphere configurations emerge from the eigenvalues problem. This requires a different regularization of the model which then allows us exact integration. 

The path integral (\ref{patheff0}) can also be rewritten as
\begin{eqnarray}
Z_{\rm eff}
&=&\int \prod_{i=1}^Nd\lambda_i \prod_{i\neq j}\frac{\lambda_i-\lambda_j}{\lambda_i-\lambda_j+1}\exp\bigg(-N t \sum_{i}\lambda_i^2\bigg).
\end{eqnarray}
By using Cauchy formula 
 this  path integral can be brought to the form 

\begin{eqnarray}
Z_{\rm eff}
&=&\sum_{\sigma\in{\cal S}_N}(-1)^{\sigma}\int\prod_{i=1}^N\bigg( d\lambda_i \frac{e^{-N t \lambda_i^2}}{\lambda_i-\lambda_{\sigma(i)}+1}\bigg).
\end{eqnarray}
It is natural to understand the integrals as contour integrals. The poles are on the real line so we regularize this partition function as follows
\begin{eqnarray}
Z_{\rm eff}
&=&\sum_{\sigma\in{\cal S}_N}(-1)^{\sigma}\oint\prod_{i=1}^N\bigg( d\lambda_i \frac{e^{-N t \lambda_i^2+i\beta\lambda_i}}{\lambda_i-\lambda_{\sigma(i)}+1+i\beta}\bigg)~,~\beta >0.
\end{eqnarray}
Thus, we close the contours in the upper half-plane. The reason behind this way of regularization is twofold. Firstly, among the two poles which may appear for each variable only one will be counted. Secondly, the result obtained here will be consistent with the result for Yang-Mills quantum mechanics obtained in \cite{Hoppe:1999xg}.

 Since the eigenvalues $\lambda_i$ are defined up to a permutation, we must also incorporate a combinatorial factor equal $1/N!$. Furthermore, the tracelessness condition $Tr\Lambda=0$ must be included in the form  $N\delta(\lambda_1+...+\lambda_N)$. In summary, we get the partition function
 \begin{eqnarray}
Z_{\rm eff}
&=&\frac{1}{(N-1)!}\sum_{\sigma\in{\cal S}_N}(-1)^{\sigma}\oint \delta(\lambda_1+...+\lambda_N)\prod_{i=1}^N\bigg( d\lambda_i \frac{e^{-N t \lambda_i^2+i\beta\lambda_i}}{\lambda_i-\lambda_{\sigma(i)}+1+i\beta}\bigg).
\end{eqnarray}
 Among the $N!$ integrals, there are only $(N-1)!$ which are non zero. They all lead to identical contributions. We get then

 \begin{eqnarray}
Z_{\rm eff}
&=&(-1)^{N-1}\oint \delta(\lambda_1+...+\lambda_N)\prod_{i=1}^N\bigg( d\lambda_i \frac{e^{-N t \lambda_i^2+i\beta\lambda_i}}{\lambda_i-\lambda_{i-1}+1+i\beta}\bigg)~,~\lambda_0\equiv\lambda_{N}.\nonumber\\
\end{eqnarray}
Next, we will perform the integrals over the variables $\lambda_2$,...,$\lambda_{N-1}$ using the residue theorem. The two remaining integrals over $\lambda_1$ and $\lambda_N$ will be constrained such that $\lambda_1+\lambda_N=0$. We introduce $i\gamma=1+i\beta$. There will be a pole in the $\lambda_2-$plane  at $\lambda_2=\lambda_3+i\gamma$, a pole in the $\lambda_3-$plane at $\lambda_3=\lambda_4+i\gamma$, a pole in the $\lambda_4-$plane at $\lambda_4=\lambda_5+i\gamma$...and a pole in the $\lambda_{N-1}-$plane  at $\lambda_{N-1}=\lambda_N+i\gamma$. The integration over $\lambda_1$ will then be seen to be dominated by the pole at $\lambda_1=\lambda_N+(N-1)i\gamma$. The delta function becomes $\delta\big(\lambda_N+(N-1)/2\big)$. In other words, we obtain, in the limit $\beta\longrightarrow 0$), the contour integral
 \begin{eqnarray}
Z_{\rm eff}
&=&-(2\pi i)^{N-1}\oint d \lambda_N \frac{1}{N^2}\delta\big(\lambda_N+\frac{N-1}{2}\big)e^{-N t \big(\lambda_N^2+(\lambda_N+(N-2))^2+...+(\lambda_N+1)^2+\lambda_N^2\big)}.\nonumber\\
\end{eqnarray}
Equivalently
\begin{eqnarray}
Z_{\rm eff}
&=&(-1)^N(2\pi i)^{N-1}\frac{1}{N^2}\exp\big(-N t\sum_{m=-\frac{N-1}{2}}^{m=\frac{N-1}{2}}m^2\big)\nonumber\\
&=&(-1)^N(2\pi i)^{N-1}\frac{1}{N^2}\exp\big(-\frac{N^2t }{3}s(s+1)\big)~,~s=\frac{N-1}{2}.
\end{eqnarray}
The smallest eigenvalue is $\lambda_N=-(N-1)/2$ while the largest eigenvalue is $\lambda_1=(N-1)/2$. We observe that $\lambda_1=\lambda_N+N-1$. We have in total $N=2s+1$ eigenvalues between $\lambda_N$ and $\lambda_1$ with a step equal $1$, viz $m=(N-1)/2,(N-3)/2,...,-(N-3)/2,-(N-1)/2$. This vacuum configuration corresponds precisely to the $SU(2)$ irreducible representation $s=(N-1)/2$.

In summary, we have found that the partition function is dominated by the integration in the vicinity of the poles $\lambda_i-\lambda_j+1=0$ which corresponds to the irreducible representation of $SU(2)$ of size $N$.

\section{Synthesis of other approaches}
The subsequent two subsections are somewhat separate, and thus can be read independently.
\subsection{Hoppe and inverted oscillator problems}
Let us again consider the path integral (with the scaling $X_i\longrightarrow \alpha D_i/\sqrt{8}$, $X_3\longrightarrow \alpha D_3/\sqrt{16}$ and $t=\alpha^4/32$)
\begin{eqnarray}
Z_{\rm CS}=\int dD_1~dD_2~dD_3 \exp N t\bigg(-2iTr D_3[D_1,D_2]-Tr D_3^2-TrD_i^2\bigg).\label{massive}
\end{eqnarray}
The action is precisely the one given in (\ref{CSfund}) after scaling, and by integrating $D_3$ we get back the effective path integral (\ref{eff1}),viz
\begin{eqnarray}
Z_{\rm eff}=\int dD_1 dD_2 \exp\bigg(-Nt Tr[D_1,D_2]^2-Nt TrD_i^2\bigg).\label{massive1}
\end{eqnarray}
This path integral (\ref{eff1}) looks unstable, and there is indeed some evidence from Monte Carlo simulations that the massive Chern-Simons theory (\ref{massive}) is ill defined. We  will elaborate on this point here shortly. 

First let us go to equation (\ref{CSfund0}) and  integrate over $X_2$, or over $X_1$, to obtain the path integral

\begin{eqnarray}
Z_{\rm eff}^{'}=\int dD_4 dD_1 \exp\bigg(Nt Tr[D_4,D_1]^2+Nt Tr D_4^2-Nt TrD_1^2\bigg).
\end{eqnarray}
We have implicitly assumed that $D_4$ is hermitian. This is still different from Hoppe's two-matrix model \cite{Hoppe:1982}, studied in \cite{Filev:2013pza,O'Connor:2012vr}, in which the matrix $D_4$ comes with a stable mass term, viz
\begin{eqnarray}
Z_{\rm Hoppe}=\int dD_4 dD_1 \exp\bigg(Nt Tr[D_4,D_1]^2-Nt Tr D_4^2-Nt TrD_1^2\bigg).
\end{eqnarray}
We write these model collectively as (in this section $g^2=1/t$)
\begin{eqnarray}
Z=\int dX dY \exp\bigg(N g^2 Tr[X,Y]^2\pm N Tr X^2-N TrY^2\bigg).
\end{eqnarray}
We would like to sketch here the approach of \cite{O'Connor:2012vr}, the results of \cite{Berenstein:2008eg} near $t=0$ (where the matrices are nearly commuting), and some of the exact results of \cite{Kazakov:1998ji}.

First we may diagonalize $Y$ to obtain 
\begin{eqnarray}
Z=\int dX d\Lambda \exp\bigg(\sum_{i<j}\ln (\lambda_i-\lambda_j)^2\bigg)\exp\bigg(\pm N \sum_{i,j}X_{ij}X_{ji}\big[1\mp g^2 (\lambda_i-\lambda_j)^2\big]\bigg)\exp\bigg(-N \sum_i \lambda_i^2\bigg).
\end{eqnarray}
For the Hoppe model (the minus sign) we can do the integral without any problem. For the other cases the integral is only formal. By performing then the integral over $X$ we can obtain an effective action in an obvious way. The approach of \cite{O'Connor:2012vr} is a generalization of this procedure. It consists of the following ingredients:
\begin{itemize}
\item{}We split the matrices $X\equiv X^1$ and $Y\equiv X^2$ to diagonal components $x_i^1$ and $x_i^2$ respectively, and off diagonal components $a_{ij}^1$ and $a_{ij}^2$ respectively.
\item{}We employ $SU(N)$ symmetry to impose the axial gauge condition
\begin{eqnarray}
\vec{n}.\vec{a}_{ij}=0.
\end{eqnarray} 
$\vec{n}$ is a constant unit vector. This condition is the analogue of diagonalizing $Y$, and as a consequence, the analogue of the Vandermonde determinant is precisely the Faddeev-Popov determinant given in this case by
 \begin{eqnarray}
S_{\rm FP}=-\frac{1}{2}\sum_{i\ne j}\log (\vec{n}.(\vec{x}_i-\vec{x}_j))^2.
\end{eqnarray} 
\item{}We integrate out the off diagonal elements. The effective action is given by the exact result
\begin{eqnarray}
S_{\rm eff}=\frac{1}{N}\sum_i(\vec{n}.\vec{x}_i)^2-\frac{1}{2N^2}\sum_{i\ne j}\log \frac{\big[\vec{n}.(\vec{x}_i-\vec{x}_j)\big]^2}{1\pm g^2\big[\vec{n}.(\vec{x}_i-\vec{x}_j)\big]^2}.
\end{eqnarray}
The plus sign corresponds to the Hoppe's model, while the minus sign corresponds to our Chern-Simons model (\ref{massive}) or equivalently (\ref{massive1}).
\end{itemize}
In the large $N$ limit we introduce, as usual, a normalized eigenvalues distribution $\rho_2(\vec{x})$. This is a rotationally invariant lifted form of the one-dimensional eigenvalues distribution of the matrix $\vec{n}.\vec{X}$ which has eigenvalues $\vec{n}.\vec{x}$. In the limit $g\longrightarrow \infty$, the model is in the commuting phase, and as a consequence, diagonalizing $\vec{n}.\vec{X}$ is equivalent to diagonalizing $X$ and $Y$, and hence, $\rho_2$ becomes the joint eigenvalues distribution of the $X^{\mu}$.  

The effective action becomes (including a Lagrange multiplier $\mu$) given by
\begin{eqnarray}
S_{\rm eff}=\int d^2x \rho_2(\vec{x})(\vec{n}.\vec{x})^2-\frac{1}{2}\int d^2x d^2x^{'}\rho_2(\vec{x})\rho_2(\vec{x}^{'})\log \frac{\big[\vec{n}.(\vec{x}-\vec{x})\big]^2}{1\pm g^2\big[\vec{n}.(\vec{x}-\vec{x})\big]^2}+\mu(\int d^2x\rho_2(\vec{x}) -1).\nonumber\\
\end{eqnarray}
By varying with respect to $\rho$ we get
\begin{eqnarray}
\mu+(\vec{n}.\vec{x})^2=\int d^2x^{'}\rho_2(\vec{x}^{'})\log \frac{\big[\vec{n}.(\vec{x}-\vec{x})\big]^2}{1\pm g^2\big[\vec{n}.(\vec{x}-\vec{x}^{'})\big]^2}.
\end{eqnarray}
We apply now the operator $\vec{n}.\vec{\nabla}_x$ to obtain
\begin{eqnarray}
\vec{n}.\vec{x}=\int d^2x^{'}\rho_2(\vec{x}^{'})\frac{1}{[\vec{n}.(\vec{x}-\vec{x}^{'})]\bigg[1\pm g^2\big[\vec{n}.(\vec{x}-\vec{x}^{'})\big]^2\bigg]}.
\end{eqnarray}
This is the analogue of the saddle point equation (\ref{sad5}). Indeed, by working in the coordinates system in which $\vec{n}.\vec{x}^{'}=x^{'1}$ we arrive at the equation (with $u=\vec{n}.\vec{x}$)
\begin{eqnarray}
u=\int dx^{'1}\rho_1({x}^{'1})\frac{1}{(u-{x}^{'1})\big[1\pm g^2\big(u-{x}^{'1}\big)^2\big]}.\label{cofi}
\end{eqnarray}
\begin{eqnarray}
\rho_1({x}^{1})=\int_{-\sqrt{R^2-(x^1)^2}}^{+\sqrt{R^2-(x^1)^2}} dx^2 \rho_2(x^1,x^2).\label{12}
\end{eqnarray}
By defining $x^{'1}=y/g$, $u=z/g$, $\rho_1(x_1^{'})=g\rho(y)$, we bring the above saddle point equation to the form (\ref{sad5}), viz
\begin{eqnarray}
\frac{z}{g^2}=\int dy \frac{\rho(y)}{(z-y)(1\pm (z-y)^2)}.\label{sad5}
\end{eqnarray}
The minus sign has been studied already in section $2.5$. Here we comment on the case of the plus sign. We go essentially through the same steps as in section $2.5$. We have then the result
\begin{eqnarray}
\frac{z}{g^2}&=&\int_{z-L}^{z+L} dx \frac{\rho(z-x)}{x(1+x^2)}\nonumber\\
&=&\rho(z)\int_{z-L}^{z+L} dx \frac{1}{x(1+x^2)}-\rho^{'}(z)\int_{z-L}^{z+L} dx \frac{1}{1+x^2}+\rho^{''}(z)\int_{z-L}^{z+L} dx \frac{x}{2(1+x^2)}+...\nonumber\\
&=&-\frac{1}{2}\rho(z)\ln(1+\frac{1}{x^2})|_{z-L}^{z+L}-\rho^{'}(z)\arctan x|_{z-L}^{z+L}+\frac{1}{4}\rho^{''}(z)\ln(1+x^2)|_{z-L}^{z+L}+...\nonumber\\
\end{eqnarray}
Again we assume a one-cut symmetric quadratic distribution $\rho(z)=a+bz^2$ in the interval $[-L,+L]$, where $a=1/(2L) -bL^2/3$, with large $L$ in the limit $g\longrightarrow \infty$. As before only the above three terms will contribute. The first term is still of order $1/L^3$, the third term is still of order $1/L$, whereas the second term becomes of order $1$. 
The saddle-point equation  to leading order in $1/L$  becomes
\begin{eqnarray}
\frac{z}{g^2}
&=&-\pi\rho^{'}(z)+...
\end{eqnarray}
We immediately obtain the parabolic distribution
\begin{eqnarray}
\rho(z)
&=&\frac{3}{4L^3}(L^2 -z^2)~,~L=(\frac{3\pi}{2})^{\frac{1}{3}}g^{\frac{2}{3}}.
\end{eqnarray}
If we go to the eigenvalues distribution $\rho_1(x_1)$ of  $x^1$, we get the distribution \cite{Berenstein:2008eg} 
\begin{eqnarray}
\rho_1(x^{1})
&=&\frac{3}{4R^3}(R^2 -(x^{1})^2)~,~R=(\frac{3\pi}{2g})^{\frac{1}{3}}.\label{parabolic}
\end{eqnarray}
The scaling behavior of $L$ here is different from the one obtained in (\ref{scaling3color}) for the three-color problem which corresponds to integrating out one of the matrices in the Chern-Simons action.

The above approach of O'Connor-Filev can also be applied directly to the analytic continuation of the Chern-Simons theory (\ref{massive}). The eigenvalues problem turns out to be precisely given by (\ref{cofi}), where the one-dimensional eigenvalues distribution $\rho_1$ is related now to the three-dimensional eigenvalues distribution $\rho_3$ by an equation similar to (\ref{12}) given by
\begin{eqnarray}
\rho_1({x}^{1})=\int_{-\sqrt{R^2-(x^1)^2}}^{\sqrt{R^2-(x^1)^2}} dx^2\int_{-\sqrt{R^2-(x^1)^2-(x^2)^2}}^{\sqrt{R^2-(x^1)^2-(x^2)^2}} dx^3 \rho_3(x^1,x^2,x^3).
\end{eqnarray}
By using rotational invariance we get
\begin{eqnarray}
\rho_1({x}^{1})&=&2\pi \int_{0}^{\sqrt{R^2-(x^1)^2}} \rho d\rho \rho_3(\sqrt{\rho^2+(x^1)^2})\nonumber\\
&=&2\pi \int_{x^1}^{R} r dr \rho_3(r).
\end{eqnarray}
By differentiating with respect to $x^1$ we get the map
\begin{eqnarray}
\rho_3(r)=-\frac{\rho_1^{'}(r)}{2\pi r}.
\end{eqnarray}
A map between the one-dimensional eigenvalues distribution $\rho_1$ and the two-dimensional eigenvalues distribution $\rho_2$ can be obtained by integrating this result over the extra coordinate \cite{O'Connor:2012vr}, viz 
\begin{eqnarray}
\rho_2(r)=-\int_{-\sqrt{R^2-r^2}}^{\sqrt{R^2-r^2}} \frac{\rho_1^{'}(\sqrt{r^2+z})}{2\pi \sqrt{r^2+z^2}}dz.
\end{eqnarray}
By substituting the parabolic distribution (\ref{parabolic}), we obtain the uniform distribution in three dimensions \cite{O'Connor:2012vr}, and the hemisphere distribution in two dimensions\cite{Berenstein:2008eg}, viz 
\begin{eqnarray}
\rho_3(r)=\frac{g}{2\pi^2}.
\end{eqnarray}
\begin{eqnarray}
\rho_2(r)=\frac{3}{2\pi R^3}\sqrt{R^2-r^2}.
\end{eqnarray}
These distributions are the leading results in the limit $g\longrightarrow \infty$. The subleading corrections were also computed in \cite{Filev:2013pza}. The radius $R$ and the eigenvalues distributions $\rho_1$ were found to be given respectively by
\begin{eqnarray}
R=(\frac{3\pi}{2g})^{\frac{1}{3}}-\frac{2\log g+\log (96\pi^4)}{6\pi g}+O(\frac{1}{g^{5/3}}).
\end{eqnarray}
\begin{eqnarray}
\rho_1(x)=\frac{g}{2\pi}\bigg[(\frac{3\pi}{2g})^{\frac{2}{3}}-x^2\bigg]+\frac{x}{2\pi^2}\log \frac{(\frac{3\pi}{2g})^{\frac{1}{3}}-x}{(\frac{3\pi}{2g})^{\frac{1}{3}}+x}+\frac{1}{2\pi^2}(\frac{3\pi}{2g})^{\frac{1}{3}}+O(\frac{\log g}{g}).
\end{eqnarray}
These results can be used, for example, to compute the value of the observable $\nu=g^2<Tr X^2/N>$. A straightforward calculation gives
 \begin{eqnarray}
\nu&=&g^2\int_{-R}^R dx x^2\rho_1(x)\nonumber\\
&=&\frac{(12\pi)^{2/3}}{20}g^{4/3}-\frac{3}{(12\pi)^{2/3}}g^{2/3}+O(g^0).
\end{eqnarray}
This is in agreement with the exact implicit result of  \cite{Kazakov:1998ji}.
\subsection{The three-color problem}
Let us now sketch how do we arrive to the eigenvalues problem (\ref{eigenvalueP}) starting from (\ref{massive}). By integrating over $D_3$ we get
\begin{eqnarray}
Z_{\rm CS}
&=&\int dD_1~dD_2 \exp Nt\bigg(-Tr[D_1,D_2]^2- TrD_i^2\bigg).
\end{eqnarray}
Next we diagonalize the hermitian matrix $D_2$. We have
\begin{eqnarray}
dD_2=\prod_{i=1}^Nd\phi_i \prod_{i<j}(\phi_i-\phi_j)^2.
\end{eqnarray}
We get then
\begin{eqnarray}
Z_{\rm CS}&=&\int \prod_{i=1}^Nd\phi_i 
\prod_{i<j}(\phi_i-\phi_j)^2\int dD_1 \exp\bigg(-\sum_{i,j}(D_1)_{ij}(D_1)^*_{ij}\bigg[-Nt(\phi_i-\phi_j)^2+Nt\bigg]-Nt\sum_{i}\phi_i^2\bigg).\nonumber\\
\end{eqnarray}
Integrating over $D_1$ we get the path integral \cite{Ydri:2012bq}

\begin{eqnarray}
Z_{\rm CS}
&=&\int \prod_{i=1}^Nd\phi_i 
\prod_{i<j}(\phi_i-\phi_j)^2 \prod_{i<j}\bigg(1-(\phi_i-\phi_j)^2\bigg)^{-1} \exp\bigg(-Nt\sum_{i}\phi_i^2\bigg).\label{patheff1}\nonumber\\
\end{eqnarray}
An almost the same eigenvalues problem can also be obtained from the following matrix model  \cite{Eynard:1998fn}
\begin{eqnarray}
Z_{\rm coloring}=\int dD_1~dD_2~dD_3 \exp Tr N\bigg(D_3\{D_1,D_2\}-\frac{1}{2}D_i^2-\frac{1}{g}V(D_3)\bigg).\label{coloring}
\end{eqnarray}
Observe that it is the anticommutator that appears in the cubic term and not the commutator. By integrating $D_2$ we get
\begin{eqnarray}
Z_{\rm coloring}=\int dD_1~dD_3 \exp Tr N\bigg(\frac{1}{2}\{D_1,D_3\}^2-\frac{1}{2}D_1^2-\frac{1}{g}V(D_3)\bigg).
\end{eqnarray}
Remark now that the "Yang-Mills term" appears as a square of an anticommutator as opposed to a commutator. Next we diagonalize the hermitian matrix $D_3$. We have
\begin{eqnarray}
dD_3=\prod_{i=1}^Nd\phi_i \prod_{i<j}(\phi_i-\phi_j)^2.
\end{eqnarray}
We get then
\begin{eqnarray}
Z_{\rm coloring}&=&\int \prod_{i=1}^Nd\phi_i 
\prod_{i<j}(\phi_i-\phi_j)^2\int dD_1 \exp\bigg(-\sum_{i,j}(D_1)_{ij}{\cal M}_{ij,kl}(\phi)(D_1)_{lk}-\frac{N}{g}V(\phi)\bigg).\nonumber\\
\end{eqnarray}
\begin{eqnarray}
{\cal M}_{ij,kl}(\phi)=\frac{N}{2}\delta_{ik}\delta_{jl}\bigg[1-(\phi_i+\phi_j)^2\bigg].
\end{eqnarray}
Integrating $D_1$ we obtain
\begin{eqnarray}
Z_{\rm coloring}&=&\int \prod_{i=1}^Nd\phi_i 
\prod_{i<j}(\phi_i-\phi_j)^2\det{\cal M}(\phi)^{-\frac{1}{2}}\exp\bigg(-\frac{N}{g}V(\phi)\bigg).\nonumber\\
\end{eqnarray}
Equivalently
\begin{eqnarray}
Z_{\rm coloring}
&=&\int \prod_{i=1}^Nd\phi_i 
\prod_{i<j}(\phi_i-\phi_j)^2 \prod_{i,j}\bigg(1-(\phi_i+\phi_j)^2\bigg)^{-\frac{1}{2}} \exp\bigg(-\frac{N}{g}V(\phi)\bigg)\label{patheff2}.\nonumber\\
\end{eqnarray}
 Let us then consider now the slightly generalized coloring model given by
\begin{eqnarray}
Z_{\rm coloring}
&=&\int \prod_{i=1}^Nd\phi_i e^{-NV_{\rm eff}(\phi)}.
\end{eqnarray}
\begin{eqnarray}
V_{\rm eff}(\phi)=\frac{1}{g}V-\frac{1}{N}\sum_{i<j}\ln (\phi_i-\phi_j)^2+\frac{n}{2N}\sum_{i,j}\ln\bigg[1-(\phi_i\pm\phi_j)^2\bigg].
\end{eqnarray}
We have considered both the plus and minus signs in the last term to cover both situations, and considered $n$ copies of the matrices $D_1$ and $D_2$. The saddle point equation reads now \footnote{In this section we have used simply $g$ instead of $g^2$, and as a consequence, we have $t=1/g$.}
\begin{eqnarray}
t\frac{\partial V}{\partial \phi_k}
&=&\frac{2}{N}\sum_{j\neq k}\frac{1}{\phi_k-\phi_j}-\frac{n}{N}\sum_{j\neq k}\bigg[\frac{1}{1+(\phi_k\pm\phi_j)}-\frac{1}{1-(\phi_k\pm \phi_j)}\bigg].
\end{eqnarray}
We introduce as usual the eigenvalue distribution $\rho$ and the resolvent $W(z)=\int dx \rho(x)/(z-x)=-W(-z)$. The loop equation turns out to be the same for both the plus  sign ((\ref{patheff2})) and the minus sign ( (\ref{patheff1})), and is given by
\begin{eqnarray}
t V^{'}(z)
&=&W(z+i0)+W(z-i0)+n\big[W(1-z)-W(1+z)\big].
\end{eqnarray}
 Our Chern-Simons model is therefore equivalent to the three-color problem. The two eigenvalues problems (\ref{patheff1}) and (\ref{patheff2}) become singular when
\begin{eqnarray}
(\phi_i\pm \phi_j)^2\longrightarrow 1.
\end{eqnarray}
In the case of (\ref{patheff2}) this corresponds to when the eigenvalues approach $\pm 1/2$. We solve thus for a cut $[a,b]$ such that $[a,b]\subset [-1/2,1/2]$. Since the model  (\ref{patheff1}) is effectively equivalent to  (\ref{patheff2}) we conclude that the Chern-Simons model is singular when the eigenvalues approach $\pm 1/2$. The eigenvalues distribution, of interest, of the Chern-Simons model is then a single cut $[a,b]\subset [-1/2,1/2]$, and $W(z)$ is analytic everywhere in the complex plane except along the cut $[a,b]$. Several remarks are in order: 
\begin{itemize}
\item{} As $z$ crosses the cut $[a,b]$ from the first sheet into the second sheet we see that the resolvent $W(z)$ becomes a linear combination of $W(z)$, $W(1-z)$ and $W(-1-z)$. In other words, we have three cuts $[a,b]$, $[1-b,1-a]$ and $[-1-b,-1-a]$ in the second sheet. By crossing the cuts $[a,b]$, $[1-b,1-a]$ and $[-1-b,-1-a]$ into the third sheet we generate more cuts. Hence the domain of definition of $W(z)$ is a Riemann surface of infinite genus with an infinite number of cuts in each sheet. 

The critical behavior when the eigenvalues approach $\pm 1/2$ corresponds to the case when all cuts merge.

By comparison the one-matrix model (we set $2 z$ to $ V^{'}(z)$ and $n[W(-1-z)+W(1-z)]$ to $0$) yields a Riemann surface with one cut in each of the two possible sheets, whereas the $O(n)$-matrix model (we set $2  z$ to $ V^{'}(z)$ and $n[W(-1-z)+W(1-z)]$ to $n W(-z)$) yields a Riemann surface with two cuts in each of the two possible sheets.

\item{} It is not difficult to show that the free theory $g=0$, for $V^{'}(z)=2z$, corresponds to a density of eigenvalues given by the Wigner's semicircle law
\begin{eqnarray}
\rho(x)=\frac{2}{\pi a^2}\sqrt{a^2-x^2}~,~a^2=2g.
\end{eqnarray}
\item{}For a quadratic potential the cut is given by $[-a,a]\subset [-1/2,1/2]$ where, from the eigenvalues problem (\ref{patheff2}),  $a\longrightarrow 0$ as $g\longrightarrow 0$. A perturbative solution around $g=0$ can be constructed for all values of $g$ less or equal than $g_c$ such that \cite{Eynard:1998fn}
\begin{eqnarray}
\frac{1}{g_c} =16+16 n(\frac{4}{\pi}-1)+O(n^2).
\end{eqnarray}
Thus the Chern-Simons model does only make sense for the values of $\alpha$ such that (with $n=1$)
\begin{eqnarray}
\alpha\geq \alpha_c\simeq \frac{8}{(2\pi)^{1/4}}+...=5.05+....
\end{eqnarray}
In other words, it seems that we can not access the important limit $\alpha\longrightarrow 0$ in this model. The above number is only a rough estimation since we do not know the terms proportional to higher powers of $n$ which can not clearly be neglected. 

There remains the possibility that the critical value $g_c$ is very large, or even infinite, in which case we may be able to take the limit $\alpha\longrightarrow 0$. This conjecture is further motivated by the fact that the original path integral with action given by the supersymmetric model (\ref{fund2}), which reduces under localization to the Chern-Simons theory (\ref{CSfund}), is perfectly well behaved in the limit $\alpha\longrightarrow 0$ \cite{Austing:2001ib}.
\end{itemize}

\section{Conclusion and Outlook}
In this article we have attempted to review, very briefly, various aspects of Yang-Mills matrix models in dimensions $D\leq 4$, which are important to the phenomena of emergent geometry and noncommutative gauge theory on the fuzzy sphere. Furthermore, we have attempted to derive the eigenvalues distribution of the $D=3$ Yang-Mills-Chern-Simons model of  Alekseev, Recknagel and Schomerus, in its matrix or Yang-Mills phase, where the matrices are nearly commuting, by escalating the problem to $4$ dimensions and including supersymmetry, then reducing the model, by means of localization, to a $D=2$ Yang-Mills matrix model (the three-color problem), which is the analytic continuation of the Hoppe's model. We have conjectured that this model is perfectly well behaved in the limit $\alpha\longrightarrow 0$, despite of the apparent instability, and derived from it the one-dimensional eigenvalues distribution, which is found to be parabolic with the correct scaling of the radius, in good accord with Monte Carlo results. Within this scheme the fuzzy sphere seems to be metastable. This remains to be re-derived and confirmed in a more rigorous way.
 
Many point require further clarification and study such as:
\begin{itemize}
\item The physics of commuting matrix models and its geometric content. Commuting matrix models are explored in \cite{Filev:2014jxa} where generalization of the hermitian quartic matrix model $V=a\phi^2+b\phi^4$, to $p$ dimensions with $SO(p)$ symmetry, is considered. The resulting actions read $V=a|\vec{\phi}|^2+b|\vec{\phi}|^4$ where $\vec{\phi}$ is an array of $p$ commuting hermitian  $N\times N$ matrices. The celebrated $3$rd order one-cut to two-cut transition gets therefore generalized to a $6$th order transition, for $p=2$, from a disk to annulus, and to a $4$th order transition, for $p=3$, from  a ball to a shell. Although these results are very exciting, it is still not clear how they actually relate to Yang-Mills matrix models.

\item The above studied Yang-Mills models with quartic terms are considered in \cite{Johnston:1997vy}, in the context of vertex models on planar graphs, where it is observed that they are equivalent to free fermion models, and the critical behavior is Ising-like since the critical exponents are found to be identical. It is very important to clarify whether these models, or variant thereof, fall indeed in the Ising universality class.

\item Another, very promising, line of investigation is the application and development of the renormalization group approach proposed by Brezin and Zinn-Justin to Yang-Mills matrix models in two dimensions with quadratic and quartic terms. This was initiated in \cite{Kawamoto:2013laa} with, very interesting, finding such as the result that the critical behavior of these models is very similar to Euclidean two-dimensional gravity.
\end{itemize}

\appendix

\section{Supersymmetry and Localization}
The primary goal in this section is to derive the Chern-Simons action (\ref{CSfund}) starting from the action (\ref{fund2}).

\paragraph{Mass Deformation:}
The reduction to one dimension of the four-dimensional ${\cal N}=1$ supersymmetric Yang-Mills theory is given by the Lagrangian 
\begin{eqnarray}
{L}_0=\frac{1}{g^2}Tr\bigg(\frac{1}{2}(D_0X_i)^2+\frac{1}{4}[X_i,X_j]^2-\frac{1}{2}\bar{\psi}{\gamma}^0D_0\psi+\frac{i}{2}\bar{\psi}{\gamma}^i[X_i,\psi] +\frac{1}{2}F^2\bigg).
\end{eqnarray}
\begin{eqnarray}
D_0={\partial}_0-i[X_0,.].
\end{eqnarray}
Let $\mu$ be a constant mass parameter. A mass deformation of the Lagrangian density ${\cal L}_0$ takes the form
\begin{eqnarray}
{L}_{\mu}={L}_0+\frac{\mu}{g^2}{L}_1+\frac{{\mu}^2}{g^2}{L}_2+...\label{4DL}
\end{eqnarray}
By dimensional analysis, the most general forms of ${\cal L}_i$  must be given by
\begin{eqnarray}
{L}_1=Tr\bigg(\bar{\psi}M{\psi}+\frac{1}{3!}S_{abc}X_aX_bX_c+J_{ab}X_aD_0X_b\bigg).
\end{eqnarray}
\begin{eqnarray}
{L}_2=Tr\bigg(-\frac{1}{2!}S_{ab}X_aX_b\bigg).
\end{eqnarray}
\begin{eqnarray}
{L}_i=0~,~i\geq 3.
\end{eqnarray}
We can follow the method of \cite{Kim:2006wg} to determine the exact form of the mass deformation. After a long calculation, we get the two-parameter  action \cite{Ydri:2012bq}

\begin{eqnarray}
{ L}_{\mu}
&=&{L}_{0}+\frac{1}{4g^2}Tr\bar{\psi}\big({\mu}_1+{\mu}_2{\gamma}^1{\gamma}^2{\gamma}^3\big){\psi}-i{\epsilon}_{ijk}\frac{{\mu}_2}{3g^2}TrX_iX_jX_k-\frac{1}{18g^2}({\mu}_1^2+{\mu}_2^2)TrX_i^2.\nonumber\\
\end{eqnarray}
This is an $SO(3)$ invariant theory. The corresponding supersymmetry transformations are given by
\begin{eqnarray}
&&{\delta}_{\mu} X_{0}=\bar{\epsilon}{\gamma}_{0}\psi\nonumber\\
&&{\delta}_{\mu} X_{i}=\bar{\epsilon}{\gamma}_{i}\psi\nonumber\\
&&{\delta}_{\mu} {\psi}=\bigg[-\frac{1}{2}[{\gamma}^{0},{\gamma}^{i}]D_0X_i+\frac{i}{4}[{\gamma}^{i},{\gamma}^{j}][X_i,X_j]-\frac{1}{3}\big({\mu}_1-{\mu}_2{\gamma}^1{\gamma}^2{\gamma}^3\big){\gamma}^iX_i\bigg]\epsilon.
\end{eqnarray}
The supersymmetry parameter $\epsilon$ is time-dependent given by  
\begin{eqnarray}
\epsilon\equiv \epsilon(t)=\exp{\frac{t}{6}\big({\mu}_1{\gamma}^0-{\mu}_2{\gamma}^0{\gamma}^1{\gamma}^2{\gamma}^3\big)}.
\end{eqnarray}
Towards a reduction to zero dimension of the four-dimensional ${\cal N}=1$ supersymmetric Yang-Mills theory, we consider now the action given by
\begin{eqnarray}
{S}_{\mu}&=&{S}_{0}+\frac{a}{4g^2}Tr\bar{\psi}\big({\mu}_1+{\mu}_2{\gamma}^1{\gamma}^2{\gamma}^3\big){\psi}-i{\epsilon}_{ijk}\frac{b{\mu}_2}{3g^2}TrX_iX_jX_k-\frac{c}{18g^2}({\mu}_1^2+{\mu}_2^2)TrX_i^2.\nonumber\\
\end{eqnarray}
\begin{eqnarray}
{S}_0=\frac{1}{g^2}Tr\bigg(\frac{1}{4}[X_{\mu},X_{\nu}][X^{\mu},X^{\nu}]+\frac{i}{2}\bar{\psi}{\gamma}^{\mu}[X_{\mu},\psi] \bigg).
\end{eqnarray}
In above, we have allowed for the possibility that mass deformations, corresponding to the zero-dimensional and one-dimensional reductions,  can be different by including different  coefficients $a$, $b$ and $c$ in front of the fermionic mass term, the Myers term, and the bosonic mass term respectively. However, we will keep the mass deformed supersymmetric transformations unchanged. After some calculation, we obtain the model 

\begin{eqnarray}
{S}_{\mu}&=&\frac{1}{g^2}Tr\bigg[\frac{1}{4}[X_{\mu},X_{\nu}][X^{\mu},X^{\nu}]+\frac{i}{2}\bar{\psi}{\gamma}^{\mu}[X_{\mu},\psi] +\frac{{\mu}_2}{6}Tr\bar{\psi}{\gamma}^1{\gamma}^2{\gamma}^3{\psi}-\frac{{\mu}_2^2}{18}TrX_i^2\nonumber\\
&-&i{\epsilon}_{ijk}\frac{{\mu}_2}{3}TrX_iX_jX_k\bigg].
\end{eqnarray}
Since $\psi$ and $\epsilon$ are Majorana spinors we can rewrite them as
\begin{eqnarray}
&&\psi=\left(
\begin{array}{c}
i{\sigma}_2({\theta}^+)^T \\
{\theta}
\end{array}
\right)~,~\epsilon=\left(
\begin{array}{c}
i{\sigma}_2({\omega}^+)^T \\
{\omega}
\end{array}
\right).
\end{eqnarray}
We compute, with  $X_0=iX_4$, the action
\begin{eqnarray}
{S}_{\mu}
&=&\frac{1}{g^2}Tr\bigg[\frac{1}{2}[X_{4},X_{i}]^2+\frac{1}{4}\bigg([X_i,X_j]-i\frac{{\mu}_2}{3}{\epsilon}_{ijk}X_k\bigg)^2+{\theta}^+\bigg(i[X_4,..]+{\sigma}_i[X_i,..]+\frac{{\mu}_2}{3}\bigg)\theta\bigg].\nonumber\\
\end{eqnarray}
The supersymmetric transformations are
\begin{eqnarray}
&&{\delta}_{\mu}X_0=i({\omega}^+\theta-{\theta}^+\omega)\nonumber\\
&&{\delta}_{\mu}X_i=i({\theta}^+{\sigma}_i{\omega}-{\omega}^+{\sigma}_i\theta)\nonumber\\
&&{\delta}_{\mu}\theta=\bigg(-i{\sigma}_i[X_0,X_i]-\frac{1}{2}{\epsilon}_{ijk}{\sigma}_k[X_i,X_j]+\frac{i}{3}{\mu}_2{\sigma}_iX_i\bigg)\omega.
\end{eqnarray}

\paragraph{Cohomological Deformation:}
The reduction to zero dimension of the four-dimensional ${\cal N}=1$ supersymmetric Yang-Mills theory is given by
\begin{eqnarray}
S_{\rm co}
&=&-\frac{1}{4}Tr[X_{\mu},X_{\nu}]^2-Tr{\theta}^+\bigg(i[X_4,..]+{\sigma}_a[X_a,..]\bigg)\theta+2Tr B^2.\label{loc2i}
\end{eqnarray}
We introduce the BRST fields 
\begin{eqnarray}
B=H+\frac{1}{2}[X_1,X_2].
\end{eqnarray}
\begin{eqnarray}
\theta_1=\eta_2+i\eta_1~,~\theta_2=\chi_1+i\chi_2.
\end{eqnarray}
\begin{eqnarray}
\phi=\frac{1}{2}(X_3+iX_4)~,~\bar{\phi}=-\frac{1}{2}(X_3-iX_4).
\end{eqnarray}
The actions  $S_{\rm co}$ becomes
\begin{eqnarray}
S_{\rm co}
&=&2Tr\big(H^2+H[X_1,X_2]+[X_i,\phi][X_i,\bar{\phi}]+[\phi,\bar{\phi}]^2-\eta_i[\phi,\eta_i]-\chi_i[\bar{\phi},\chi_i]\nonumber\\
&-&\eta_1\epsilon^{ij}[{\chi}_i,X_j]+\eta_2[\chi_i,X_i]\big).
\end{eqnarray}
This action has  four independent real supersymmetries. It can be cohomologically deformed along the lines of  \cite{Austing:2001ib}. By requiring $SO(3)$ covariance, a Myers term, and mass terms for all the bosonic and fermionic matrices we arrive at the three-parameter cohomological deformation \cite{Ydri:2012bq}
\begin{eqnarray}
S_{\rm def}=S_{\rm co}+\hat{S}+\kappa_2\Delta \hat{S}.
\end{eqnarray}
The actions $\hat{S}$ and $\Delta\hat{S}$ are given by 
\begin{eqnarray}
\hat{S}&=&2i\kappa_1Tr(\chi_1\chi_2-\eta_1\eta_2)+\epsilon\kappa_1TrX_i^2+2i\kappa_1Tr\bar{\phi}H-2i(4\epsilon-\kappa_1)Tr\phi H-4\epsilon(2\epsilon-\kappa_1)Tr\phi^2\nonumber\\
&-&8i\epsilon Tr\phi[X_1,X_2]+2i(\kappa_1+2\epsilon)Tr\bar{\phi}[X_1,X_2].
\end{eqnarray}
\begin{eqnarray}
\Delta\hat{S}&=&2iTr\eta_1\eta_2+2iTr\bar{\phi}H+2(4\epsilon-\kappa_1)Tr\bar{\phi}\phi-2\kappa_1Tr\bar{\phi}^2.
\end{eqnarray}
The first supercharge of this cohomologically deformed theory corresponds to the supersymmetry transformations
\begin{eqnarray}
d_{\rm def} X_i=\chi_i.
\end{eqnarray}
\begin{eqnarray}
d_{\rm def}\phi=0~,~d_{\rm def}\bar{\phi}=-\eta_2.\label{susycoh1}
\end{eqnarray}
\begin{eqnarray}
  d_{\rm def}H&=&[\phi,\eta_1]+i\kappa_1\eta_2.\label{susycoh2}
\end{eqnarray}
\begin{eqnarray}
  d_{\rm def} \eta_1=H+(-i(4\epsilon-\kappa_1)\phi+i\kappa_1\bar{\phi})~,~d_{\rm def}\eta_2=[\bar{\phi},\phi].\label{susycoh3}
\end{eqnarray}
\begin{eqnarray}
  d_{\rm def}\chi_i&=&[\phi,X_i]+i\epsilon \epsilon_{ij}X_j.\label{susycoh4}
\end{eqnarray}
The theory $S_{\rm def}$ has only two independent real supersymmetries. The second supercharge can be obtained by an appropriate permutation of the spinors $\eta_i$ and $\chi_i$ \cite{Austing:2001ib}. In the limit of zero deformation,  ${S}_{\rm def}\longrightarrow S_{\rm co}$, i.e. $\kappa_1,\kappa_2,\epsilon\longrightarrow 0$, and $d_{\rm def}\longrightarrow d$. In this limit 
\begin{eqnarray}
S_{\rm co}=dTrQ\Rightarrow dS_{\rm co}=d^2TrQ=0,
\end{eqnarray}
where
\begin{eqnarray}
Q=2\bigg(-\chi_i[X_i,\bar{\phi}]+\eta_1[X_1,X_2]+\eta_1H-\eta_2[\phi,\bar{\phi}]\bigg).\label{Q}
\end{eqnarray}
Thus the exterior derivative $d$ is nilpotent on gauge invariant quantities. Similar considerations hold for the deformed action  $S_{\rm def}$ and the deformed exterior derivative $d_{\rm def}$ \cite{Austing:2001ib,Ydri:2012bq}. See also \cite{Kazakov:1998ji}.

The path integral, for $\kappa_2=0$, is given by
\begin{eqnarray}
Z_0[\kappa_1,\epsilon]&=&\int dX_1~dX_2~d\bar{\phi}~d\phi~dH~d\chi_1~d\chi_2~d\eta_1~d\eta_2~\exp(-S_{\rm co}-\hat{S}).
\end{eqnarray}
This can be rewritten as
\begin{eqnarray}
Z_0[\kappa_1,\epsilon]&=&\int dX_1~dX_2~d\bar{\phi}~d\phi~dH~d\chi_1~d\chi_2~d\eta_1~d\eta_2~\exp\bigg(-S_{\rm co}-\Delta{S}_{\rm co}+2Tr(H^2+H[X_1,X_2])\nonumber\\
&+&\frac{1}{2}Tr[X_1,X_2]^2-2\big(H+\frac{1}{2}[X_1,X_2]+\frac{i}{2}\kappa_1\bar{\phi}-\frac{i}{2}(4\epsilon-\kappa_1)\phi\big)^2\bigg).
\end{eqnarray}
The above path integral is  effectively equivalent to
\begin{eqnarray}
Z_0[\kappa_1,\epsilon]&=&\int dX_{\mu}~dB~d\theta^+~d\theta~\exp\big(-S_{\rm co}-\Delta{S}_{\rm co}\big).
\end{eqnarray}
The action $\Delta{S}_{\rm co}$ is given by
\begin{eqnarray}
\Delta{S}_{\rm co}
&=&\kappa_1Tr\theta^+\theta+\epsilon\kappa_1Tr X_a^2+\frac{1}{2}\kappa_1(2\epsilon-\kappa_1)TrX_4^2-\frac{i}{3}(4\epsilon+\kappa_1)\epsilon_{abc}TrX_aX_bX_c.\nonumber\\
\end{eqnarray}
We have the full action
\begin{eqnarray}
S_2[\kappa_1,\epsilon]=\Delta{S}_{\rm co}+\Delta{S}_{\rm co}.
\end{eqnarray}
We note that the action $S_{2}[\kappa_1,\epsilon]$, for the values $\kappa_1=2\epsilon$, $\kappa_2=0$, corresponds precisely to the mass deformed action considered in the previous paragraph with $\mu_2=-3\kappa_1$ and $g^2=-1$, and as a consequence, the above supersymmetry transformations (\ref{susycoh1})-(\ref{susycoh4}) will correspond to one of the mass deformed supercharges.  We have then the result
\begin{eqnarray}
S_2[\kappa_1,\epsilon=\frac{\kappa_1}{2}]=-g^2S_{\mu}|_{\mu_2=-3\kappa_1}.
\end{eqnarray}

\paragraph{Localization:}

The above theory depends a priori on  three parameters $\kappa_1$, $\kappa_2$ and $\epsilon$. We note that $\kappa_2$ does not appear explicitly in the supersymmetry transformations. As it turns out, the path integral will also not depend explicitly on $\kappa_2$. The proof goes as follows  \cite{Austing:2001ib}. The path integral we wish to evaluate is
\begin{eqnarray}
Z_{\kappa_2}[\kappa_1,\epsilon]&=&\int dX_1~dX_2~d\bar{\phi}~d\phi~dH~d\chi_1~d\chi_2~d\eta_1~d\eta_2~\exp(-S_{\rm def}).
\end{eqnarray}
By using supersymmetry transformations (\ref{susycoh1})-(\ref{susycoh4}), we can show immediately that $2id_{\rm def}(\eta_1\bar{\phi})=2id_{\rm def}\eta_1.\bar{\phi}-2i\eta_1d_{\rm def}\bar{\phi}=\Delta\hat{S}$. Hence
\begin{eqnarray}
\frac{\partial}{\partial \kappa_2}Z_{\kappa_2}[\kappa_1,\epsilon]
&=&-2i\int dX_1~dX_2~d\bar{\phi}~d\phi~dH~d\chi_1~d\chi_2~d\eta_1~d\eta_2~d_{\rm def}\big(\eta_1\bar{\phi}~\exp(-S_{\rm def})\big).\nonumber\\
\end{eqnarray}
In the above equation we have also used the  fact that $S_{\rm def}$ is invariant under the supersymmetry transformations  (\ref{susycoh1})-(\ref{susycoh4}), viz $d_{\rm def}S_{\rm def}=0$. 

Let us denote collectively the bosonic and fermionic matrices  by $A^a$. Also, let us observe, from the supersymmetry transformations  (\ref{susycoh1})-(\ref{susycoh4}), that each supersymmetric variation $d_{\rm def}A^a$ is independent of $A^a$. Then
\begin{eqnarray}
d_{\rm def}(..)=d_{\rm def}A^a\frac{\partial}{\partial A^a}(..)=\frac{\partial}{\partial A^a}(d_{\rm def}A^a..)
\end{eqnarray}
Hence, we have
\begin{eqnarray}
\frac{\partial}{\partial \kappa_2}Z_{\kappa_2}[\kappa_1,\epsilon]
&=&-2i\int dA ~\frac{\partial}{\partial A^a}\big(d_{\rm def}A^a\eta_1\bar{\phi}~\exp(-S_{\rm def})\big).
\end{eqnarray}
This is obviously $0$. We get then the interesting result that the path integral $Z_{\kappa_2}[\kappa_1,\epsilon]$ is in fact independent of $\kappa_2$. In particular, we have
\begin{eqnarray}
Z_{0}[\kappa_1,\epsilon]={\rm lim}_{\kappa_2\longrightarrow \infty}\int dX_1~dX_2~d\bar{\phi}~d\phi~dH~d\chi_1~d\chi_2~d\eta_1~d\eta_2~\exp(-S_{\rm co}-\hat{S}-\kappa_2\Delta \hat{S}).\nonumber\\
\end{eqnarray}
As we will see now, the path integral localizes in the limit $\kappa_2\longrightarrow\infty$. Indeed, we will be able to integrate the BRST quartet $\eta_1$, $\eta_2$, $H$ and $\bar{\phi}$ explicitly in this limit. 

The fermionic part of the action which depends on $\kappa_2$ is $2i\kappa_2Tr\eta_1\eta_2$. Thus, by using the saddle point method, we can see that the action in the directions of $\eta_1$ and $\eta_2$ is localized around the saddle points $\eta_1=0$ and $\eta_2=0$.

Next, we do the path integral over $H$, then over $X_3$, in that order, using again the saddle point method. The relevant terms are the bosonic contributions which are proportional to $\kappa_2$. 
For $\epsilon<0$, we can verify that the integral over $X_3$ is exponentially damped and therefore we can shift $X_3$ appropriately. The resulting integral over $H$ turns out also to be damped exponentially. Explicitly we have
\begin{eqnarray}
2(4\epsilon-\kappa_1)\kappa_2Tr\bar{\phi}{\phi}-2\kappa_1\kappa_2Tr\bar{\phi}^2+2i\kappa_2Tr\bar{\phi}H=
&-&2\epsilon\kappa_2Tr\big(X_3+\frac{i}{4\epsilon}(H-\kappa_1X_4)\big)^2\nonumber\\
&-&\frac{\kappa_2}{8\epsilon}Tr(H+(4\epsilon-\kappa_1)X_4)^2\nonumber\\
&+&\kappa_2(4\epsilon-\kappa_1)TrX_4^2.
\end{eqnarray}
In the limit $\kappa_2\longrightarrow \infty$ the hermitian matrix $H$ will be localized around $-(4\epsilon-\kappa_1)X_4=i(4\epsilon-\kappa_1)(\phi+\bar{\phi})$. We can then shift the integral over $X_3$ as $X_3\longrightarrow \bar{\phi}=-\frac{1}{2}(X_3-iX_4)$, i.e. we can  assume that   $\bar{\phi}$ is hermitian, and for consistency we will also shift the integral over $X_4$ as $X_4\longrightarrow \phi=iX_4-\bar{\phi}$. The above equation reduces to
\begin{eqnarray}
2(4\epsilon-\kappa_1)\kappa_2Tr\bar{\phi}{\phi}-2\kappa_1\kappa_2Tr\bar{\phi}^2+2i\kappa_2Tr\bar{\phi}H&=&
-8\epsilon\kappa_2Tr\bar{\phi}^2-\kappa_2(4\epsilon-\kappa_1)Tr(\phi+\bar{\phi})^2.\nonumber\\
\end{eqnarray}
Thus in the limit $\kappa_2\longrightarrow \infty$, the hermitian matrix $\bar{\phi}$ is localized around $0$. The matrix $\phi$ is then seen to be antihermitian identified with $iX_4$. Equivalently we can assume that $\phi$ is a hermitian matrix identified with $X_3$, since the saddle point in the direction $\bar{\phi}$ is $\bar{\phi}=0$.
In summary, we get by using the saddle point method the result
\begin{eqnarray}
\int d\bar{\phi}~dH~ f(\bar{\phi},H)~e^{-2(4\epsilon-\kappa_1)\kappa_2Tr\bar{\phi}{\phi}+2\kappa_1\kappa_2Tr\bar{\phi}^2-2i\kappa_2Tr\bar{\phi}H}&\sim& f(0,i(4\epsilon-\kappa_1)\phi)e^{\kappa_2(4\epsilon-\kappa_1)Tr\phi^2}.\nonumber\\
\end{eqnarray}
 The $\kappa_2$ dependence cancels completely if we choose $4\epsilon=\kappa_1$. 

The integration over the remaining fermionic degrees of freedom $\chi_1$ and $\chi_2$ is now trivial since they are free degrees of freedom decoupled from everything else. We end up with the model
\begin{eqnarray}
Z_{0}[\kappa_1,\epsilon]&=&\int d{\phi}~dX_1~dX_2 \exp\bigg(2i\kappa_1Tr\phi[X_1,X_2]-4\epsilon(\kappa_1-2\epsilon)Tr\phi^2-\epsilon\kappa_1TrX_i^2\nonumber\\
&+&\kappa_2(4\epsilon-\kappa_1)Tr\phi^2\bigg).
\end{eqnarray}
This is essentially the path integral of two-dimensional gauge theory on the fuzzy sphere studied in \cite{Ishiki:2008vf}. As was shown in \cite{Ishiki:2009vr} it can also be derived from the reduction to zero dimension of Chern-Simons theory on ${\bf S}^3$ . 


\paragraph{Summary:} 
The deformed supersymmetric Yang-Mills (YM) matrix models we have studied in \cite{Ydri:2012bq} are:
\begin{itemize} 
\item{}The mass deformed supersymmetric YM matrix model which corresponds to the cohomologically deformed action with the values $\kappa_1=2\epsilon$ and $\kappa_2=0$. This is a one-parameter matrix model given by the action

\begin{eqnarray}
S_2&=&-\frac{1}{4}Tr[X_{\mu},X_{\nu}]^2-i\kappa_1 \epsilon_{abc}TrX_aX_bX_c+\frac{\kappa_1^2}{2}Tr X_a^2\nonumber\\
&-&Tr{\theta}^+\bigg(i[X_4,..]+{\sigma}_a[X_a,..]-\kappa_1\bigg)\theta.
\end{eqnarray}
This enjoys full ${\cal N}=1$ mass deformed supersymmetry. However, the corresponding supersymmetric path integral is found (conjectured) to be not convergent for generic values of the fermion mass term \cite{Ydri:2012bq}. Hence, in  Monte Carlo simulations, we need to regularize this theory. This can be achieved, for example, by setting the mass term to zero which breaks supersymmetry explicitly.
 
\item{}The cohomologically deformed supersymmetric YM matrix model corresponding to the values $\kappa_1=0$ and $\kappa_2=0$.  This is   YM matrix model minimally deformed, i.e. with only the Chern-Simons term as a deformation, given by the one-parameter action

\begin{eqnarray}
S_2&=&-\frac{1}{4}Tr[X_{\mu},X_{\nu}]^2-\frac{4i\epsilon}{3} \epsilon_{abc}TrX_aX_bX_c-Tr{\theta}^+\bigg(i[X_4,..]+{\sigma}_a[X_a,..]\bigg)\theta.\nonumber\\
\end{eqnarray}
 This enjoys only half ${\cal N}=1$ cohomologically deformed supersymmetry. The corresponding path integral is  well defined. Indeed, it is found in \cite{Austing:2001pk,Austing:2003kd}, that  supersymmetric YM path integral in $D=4$ is always convergent, even when the Chern-Simons term is included. This case provides, therefore, an example of a non-perturbative regularization of exact supersymmetry, based on matrix models as opposed to lattice models,  which can be accessed exactly in Monte Carlo simulations.

\item{}The cohomological deformation of $4-$dimensional Yang-Mills action, for $\kappa_1=4\epsilon$, is a one-parameter matrix model given by the action
\begin{eqnarray}
S_2&=&-\frac{1}{4}Tr[X_{\mu},X_{\nu}]^2-\frac{2i\kappa_1}{3}\epsilon_{abc}TrX_aX_bX_c+\frac{\kappa_1^2}{4}Tr X_a^2-\frac{\kappa_1^2}{4}TrX_4^2\nonumber\\
&-&Tr{\theta}^+\bigg(i[X_4,..]+{\sigma}_a[X_a,..]-\kappa_1\bigg)\theta.\label{fund2F1}
\end{eqnarray}
Again, this enjoys only half ${\cal N}=1$ cohomologically deformed supersymmetry.  Also, there is here the problematic fermion mass term as well as a negative mass for the matrix $X_4$.

By employing supersymmetry and localization technique  we have shown that this theory, with  $\kappa_1 <0$ , is equivalent to the Chern-Simons matrix model
\begin{eqnarray}
S_{\rm CS}=-2i\kappa_1TrX_3[X_1,X_2]+\frac{\kappa_1^2}{2}TrX_3^2+\frac{\kappa_1^2}{4}TrX_i^2.\label{fund2F2}
\end{eqnarray}
\end{itemize}

\paragraph{Acknowledgment:} I would like to thank D.~O'Connor and V.~G.~Filev  for useful discussions during a visit to DIAS while this work was completed. 
This research was supported by CNEPRU: "The National (Algerian) Commission for the Evaluation of
University Research Projects"  under contract number ${\rm DO} 11 20 13 00 09$.

\end{document}